\documentclass{rstransa}

\begin{document}


\title{Physics-inspired analysis of the two-class income distribution in the USA in 1983--2018}


\author{
Danial Ludwig and Victor M. Yakovenko}

\address{JQI, Department of Physics, University of Maryland, College Park, Maryland, 20742, USA}

\subject{Complexity, Statistical physics, Statistics}


\corres{Victor M. Yakovenko\\
\email{yakovenk@umd.edu}}

\begin{abstract}
The first part of this paper is a brief survey of the approaches to economic inequality based on ideas from statistical physics and kinetic theory.  These include the Boltzmann kinetic equation, the time-reversal symmetry, the ergodicity hypothesis, entropy maximization, and the Fokker-Planck equation.  The origins of the exponential Boltzmann-Gibbs distribution and the Pareto power law are discussed in relation to additive and multiplicative stochastic processes.  The second part of the paper analyzes income distribution data in the USA for the time period 1983--2018 using a two-class decomposition.  We present overwhelming evidence that the lower class (more than 90\% of the population) is described by the exponential distribution, whereas the upper class (about 4\% of the population in 2018) by the power law.  We show that the significant growth of inequality during this time period is due to the sharp increase in the upper-class income share, whereas relative inequality within the lower class remains constant.  We speculate that the expansion of the upper-class population and income shares may be due to increasing digitization and non-locality of the economy in the last 40 years.

\end{abstract}


\begin{fmtext}


arXiv:2110.03140v2, 5 January 2022.  Submitted to \textit{Philosophical Transactions of the Royal Society A} for the special issue ``Kinetic exchange models of societies and economies''.

\end{fmtext}


\maketitle

\section{Introduction}

Inequality is an important, ubiquitous feature of human society.  What is the origin of economic inequality?  Is it inevitable or avoidable?  Are there common mathematical patterns of inequality?  Is quantitative description of such patterns possible?  These questions have puzzled scientists for centuries.  Lately, in the last 20 years or so, the subject has moved to the center of public discourse.

Statistical physics and kinetic theory deal with big statistical ensembles and, as such, are branches of the applied theory of probabilities.  The econophysics movement, emerging about 25 years ago \cite{Stanley-1996,Chakrabarti-history}, utilizes mathematical methods developed in these disciplines for probabilistic description of social ensembles.  Several papers in the late 1990s -- early 2000s applied these ideas to economic inequality \cite{Redner-1998,Yakovenko-2000,Chakraborti-2000,Bouchaud-2000}.  Since then, the volume of such literature has grown enormously \cite{Yakovenko-2009,Cockshott-2009,Chakrabarti-2013,Ribeiro-2020}.  Importantly, the econophysics approach has attracted attention of economists and social scientists  \cite{Rosser-2016,Shaikh-2017} and is now debated within these disciplines \cite{Scharfenaker-2020}, as well as in popular media \cite{Cho-2014,Gronewold-2021}.
Physics-inspired approaches to economic inequality are briefly reviewed  in Sec.~\ref{sec:overview} of this paper, particularly for the benefit of non-physicists.

Then, in Sec.~\ref{sec:empirics}, these ideas are applied to a quantitative analysis of the income distribution data in the United States for 1983--2018.  We use the publicly available data from Publication 1304 by the Statistics of Income research division of the Internal Revenue Service (IRS), the United States tax agency \cite{IRS-1304}.  The earliest year covered by Publication 1304 is 1983, and the latest year currently available is 2018, while the earlier data are available in Publication 79.  Our quantitative study of such a wide historical range reveals salient trends of inequality evolution in the United States.  The paper ends with Conclusions in Sec.~\ref{sec:conclusions}.

\section{Physics-inspired approaches to economic inequality}  
\label{sec:overview}

This Section briefly outlines mathematical approaches to economic inequality introduced in the early econophysics papers \cite{Redner-1998,Yakovenko-2000,Chakraborti-2000,Bouchaud-2000} and subsequently followed by numerous further extensions.  It is not intended as a comprehensive literature survey, but rather as a quick summary of the ideas relevant to our quantitative analysis of income distribution data presented in Sec.~\ref{sec:empirics}.

\subsection{The Boltzmann kinetic equation}

One approach is based on the Boltzmann kinetic equation.  First, let us consider an ideal gas, where molecules have kinetic energies $\varepsilon$ and experiencing pairwise collisions.  The total energy of two molecules remains constant before and after collision \begin{equation} \label{E1+E2}
  \varepsilon_1+\varepsilon_2=\varepsilon_1'+\varepsilon_2',
\end{equation}
while the energy $\varepsilon_1-\varepsilon_1'$ is transferred from the first to the second molecule.

By analogy, now let us consider a monetary transaction between two economic agents, whose money balances are $m_1$ and $m_2$ \cite{Yakovenko-2000}.  Upon transaction, the first agent transfers some amount of money $\Delta$ to the second agent in payment for goods or services.  Then the money balances of the agents become $m_1'=m_1-\Delta$ and $m_2'=m_2+\Delta$, so the sum is conserved
\begin{equation} \label{m1+m2}
  m_1+m_2=m_1'+m_2'.
\end{equation}
The conservation of money expressed by Eq.~(\ref{m1+m2}) is nothing but the law of accounting \cite{Yakovenko-2016}.  From the accounting perspective, all monetary transactions are transfers of digits from one account to another, so the decrease in one account is equal to the increase in another.  The flow of monetary digits, i.e.,\ the abstract bits of information, between different accounts constitutes the informational layer of the economy \cite{Yakovenko-2016}.  It is coupled to another, physical layer of the economy, which represents counterflow of physical goods and services in transactions between agents.  In contrast to the monetary digital layer, the items in the physical layer are not conserved, because they can be produced, consumed, and destroyed.  However, production of physical items, by itself, does not change the number of digital monetary units in the system, because they belong to different layers and are governed by different rules.

For a statistical ensemble of many interacting economic agents, we introduce the probability distribution $P(m,t)$ of money $m$ among the agents.  Its evolution as a function of time $t$, as a result of monetary transactions, can be described by the Boltzmann kinetic equation \cite{Landau-10}
\begin{align} \label{Boltzmann}
  \frac{\partial P(m,t)}{\partial t}= \int\!\!\!\!\int dm'\, d\Delta\, 
  & \left\{ -\Gamma_{[m,m']\to[m-\Delta,m'+\Delta]}P(m,t)P(m',t)  \right.
  \\
  & \left. +\Gamma_{[m-\Delta,m'+\Delta]\to[m,m']}
  P(m-\Delta,t)P(m'+\Delta,t) \right\}.
\nonumber
\end{align}
Here $\Gamma_{[m,m']\to[m-\Delta,m'+\Delta]}$ is the probability of transferring money $\Delta$ from an agent with money $m$ to an agent with money $m'$ per unit time.  This probability, multiplied by the occupation numbers $P(m,t)$ and $P(m',t)$, gives the rate of transitions from the state $[m,m']$ to the state $[m-\Delta,m'+\Delta]$.  The first term in Eq.\ (\ref{Boltzmann}) gives the depopulation rate of the state $m$, whereas the second term represents the reversed process, where the occupation number $P(m,t)$ increases.

In order to make further progress, different papers make various assumptions about the structure of the transition probabilities $\Gamma_{[m,m']\to[m-\Delta,m'+\Delta]}$.  For example, Ref.\ \cite{Chakraborti-2000} introduced a model of saving propensity, where the agents save some fraction of their money balances before engaging in monetary transactions described by Eq.~(\ref{m1+m2}).  Whether such specific model assumptions are realistic remains an open question.

Many papers, particularly in this volume, apply Eq.\ (\ref{Boltzmann}) not to money $m$, but to assets \cite{Redner-1998} or wealth $w$.  Some papers seem to use $m$ and $w$ synonymously, in which case there is no difference.  But more generally, wealth $w=m+ps$ is a sum of the money balance $m$ and the monetary value $ps$ of tangible property, e.g.,\ stocks $s$, where $p$ is the stock price.  Then, wealth is generally not conserved, because it may increase or decrease due to a change in price $p$, while an agent is not conducting any transactions at all.  For example, during the early COVID crisis in March 2020, many agents stayed passive and performed no transactions, but trillions of dollars of their wealth first disappeared and then reappeared due to changes of stock prices.  In contrast, money balances were largely unaffected by such wild swings.

\subsection{The equilibrium Boltzmann-Gibbs distribution}

Interestingly, a stationary solution of Eq.\ (\ref{Boltzmann}) can be found without detailed knowledge of the transition probabilities $\Gamma_{[m,m']\to[m-\Delta,m'+\Delta]}$, if they satisfy the time-reversal symmetry
\begin{equation}  \label{reversal}
  \Gamma_{[m,m']\to[m-\Delta,m'+\Delta]}=\Gamma_{[m-\Delta,m'+\Delta]\to[m,m']}.
\end{equation}
The condition (\ref{reversal}) means that the probabilities of transactions in direct and reversed directions are equal.  The fundamental dynamical equations in physics do satisfy the time-reversal symmetry.  In contrast, there is no such general requirement in economics, e.g.,\ the model \cite{Chakraborti-2000} with saving propensity does not have the time-reversal symmetry.  Nevertheless, such symmetry may be approximately applicable in some cases. 

In Eq.\ (\ref{Boltzmann}), the probability distribution $P(m)$ is stationary, $dP(m)/dt=0$, when the rates of direct and reversed transitions are equal, so that 
the two terms in the curly brackets cancel each other.  This is called the principle of detailed balance.  If the time-reversal condition (\ref{reversal}) is satisfied, then the transition probabilities $\Gamma$ cancel out, and the detailed balance condition becomes
\begin{equation}  \label{balance}
   P(m)P(m')=P(m-\Delta)P(m'+\Delta).
\end{equation}
In the notation of Eq.~(\ref{m1+m2}), Eq.~(\ref{balance}) has the form 
$P(m_1)P(m_2)=P(m_1')P(m_2')$ and can summarized as follows.  The probability $P$ is multiplicative, whereas the energy $\varepsilon$ or the money $m$ appearing in the argument of $P$ is additive, due to conservation laws (\ref{E1+E2}) or (\ref{m1+m2}).  A general solution of Eq.\ (\ref{balance}) is given by the exponential Boltzmann-Gibbs distribution \cite{Yakovenko-2000}
\begin{equation}  \label{Gibbs}
  P(m) = c\, e^{-m/T_m}.
\end{equation}
Here the coefficient $c=1/T_m$ is obtained from the normalization condition $\int_0^\infty dm\, P(m)=1$, whereas the money temperature $T_m=\langle m\rangle=\int_0^\infty dm\,P(m)\,m$ is equal to the average money balance of an agent.  

We assume that the boundary condition $m\geq0$ is imposed, so that debt, which can be interpreted as a negative money balance, is not permitted.  Thus, $m=0$ represents the lowest possible balance of an agent, analogous to the ground-state energy in physics.  Permitting debt, i.e.,\ negative $m$, makes the problem more complicated and potentially unstable \cite{Yakovenko-2016}, unless a limit on debt is imposed \cite{Yakovenko-2000,Xi-Ding-Wang-2005}

The Boltzmann-Gibbs distribution of energy $P(\varepsilon)\propto e^{-\varepsilon/T_\varepsilon}$, analogous to Eq.~(\ref{Gibbs}), plays a fundamental role in the equilibrium statistical physics.  In textbooks on this subject \cite{Landau-5}, it is usually derived using other methods, without reference to the dynamical kinetic equation (\ref{Boltzmann}).  In particular, Eq.~(\ref{Gibbs}) can be derived geometrically from the ergodicity hypothesis \cite{Lopez-Ruiz-2008,Biro-2014}, which postulates that all microscopic configurations of the system consistent with macroscopic constraints are equally probable.  

Alternatively, the Boltzmann-Gibbs distribution (\ref{Gibbs}) can be obtained by maximizing the entropy of the system $S=-\int_0^\infty dm\,P(m)\,\ln[P(m)]$ with a constraint on the total money in the system $\langle m\rangle=\int_0^\infty dm\,P(m)\,m$ using the method of Lagrange multipliers.  The entropy $S=\ln W$ is the logarithm of the multiplicity $W$, which is the number of microscopic realizations of a given distribution $P(m)$ obtained by combinatorial permutations of the agents.  The principle of entropy maximization follows from the H-theorem proved by Boltzmann \cite{Landau-10}.  The theorem shows that, for the probability distribution $P(m,t)$ obeying the Boltzmann kinetic equation (\ref{Boltzmann}), the entropy $S(t)$ monotonously increases in time, until it reaches its maximal value at the equilibrium Boltzmann-Gibbs distribution (\ref{Gibbs}).  Time evolution of global inequality in energy consumption and $\rm CO_2$ emissions per capita was interpreted in Refs.~\cite{Lawrence-2013,Semieniuk-2020} as a manifestation of entropy increase due to economic globalization.

\subsection{The Fokker-Planck equation}

When the change $\Delta$ is small for an agent with money balance $m$, the integro-differential Boltzmann kinetic equation (\ref{Boltzmann}) can be reduced the partial-differential Fokker-Planck equation \cite{Landau-10}, also known as the diffusion equation or the Kolmogorov forward equation,
\begin{equation}  \label{diffusion}
   \frac{\partial P(m,t)}{\partial t}=\frac{\partial}{\partial m} \left[A(m)P(m,t)
   + \frac{\partial}{\partial m} B(m)P(m,t) \right], 
   \quad A=-{\langle\Delta\rangle \over dt}, 
   \quad B={\langle\Delta^2\rangle \over 2 dt}.
\end{equation}
The coefficients $A$ and $B$, known as the drift and diffusion, are the first and second moments of the random money balance changes $\Delta$ per time increment $dt$.  

Assuming the boundary condition $m\geq0$, a stable stationary solution of Eq.\ (\ref{diffusion}) exists only for the drift in negative direction, $\langle\Delta\rangle/dt<0$, so that the coefficient $A>0$ is positive.  The stationary solution $\partial_tP=0$ of Eq.\ (\ref{diffusion}) is then obtained by setting the probability flux, which is the expression in the square brackets, to zero:
  \begin{equation} \label{stationary}
  P(m)=\frac{c}{B(m)}\exp\left(-\int_0^m\frac{A(m')}{B(m')}\,dm'\right),
  \end{equation}
where the coefficient $c$ follows from the normalization condition $\int_0^{\infty}P(m)\,dm=1$. 

In the simplest case where the coefficients $A$ and $B$ are constants independent of $m$, the stationary distribution (\ref{stationary}) is exponential $P(m)\propto e^{-m/T_m}$, thus reproducing the Boltzmann-Gibbs formula (\ref{Gibbs}).  Here the money temperature $T_m=B/A$ is expressed in terms of $B$ and $A$ similarly to the Einstein relation between temperature, diffusion, and mobility \cite{Landau-10}.  A well-known example in physics is the barometric distribution of gas density $P(z)\propto e^{-\mu gz/T}$ versus the height $z$ in the presence of gravity, where $\mu$ is the mass of a molecule, $g$ is the gravitational acceleration, and $\mu gz$ is the gravitational energy of a molecule.  The competition between downward drift due to gravity and spreading around due to diffusion produces the stationary exponential distribution $P(z)$.  It represents a statistical equilibrium, in contrast to a mechanical one where two forces acting on an object balance each other (like supply and demand in economics).

\subsection{Applications to income distribution}

The Boltzmann kinetic equation (\ref{Boltzmann}) is a two-body equation, in the sense that it involves the probability distributions $P(m,t)$ and $P(m',t)$ of two interacting agents, so it is a nonlinear equation.  In comparison, the Fokker-Planck equation (\ref{diffusion}) is a one-body equation, because it only involves $P(m,t)$ for one agent, so this equation is linear and sometimes called the master equation.  Coupling between agents is implicitly represented in Eq.~(\ref{diffusion}) by the first and second moments of the fluctuations $\Delta$ of the variable $m$ for a given agent, due to its interaction with the ``environment'' consisting of the other agents.   Unlike the two-body approach, the one-body description does not invoke a conservation law explicitly.  Thus, Eq.~(\ref{diffusion}) can be generally applied to all kinds of stochastic variables, which is widely done in the literature.  For example, the famous Black-Scholes equation for random fluctuations of stock prices is based on a version of Eq.~(\ref{diffusion}).  A particular version of Eq.~(\ref{diffusion}) was proposed in Ref.~\cite{Bouchaud-2000} for the distribution of wealth $w$.

In the rest of the paper, we switch to the probability distribution $P(r)$ of income $r$, instead of the money balance $m$.  The income $r$ of an agent is the influx of money per unit time, typically per year, so it is analogous to the power in physics, if money $m$ is analogous to the energy.  Following this definition, the income is non-negative, so it satisfies the boundary condition $r\geq0$.  The main reason for this change of focus is that empirical data for income distribution $P(r)$ are available from government  agencies, whereas empirical data for $P(m)$ are difficult to obtain.  Income inequality is widely discussed in the literature and is an important subject by itself.

Let us treat the income $r$ of an agent as a stochastic variable experiencing random fluctuations $\Delta$ over a time increment $dt$.  Then the time-dependent probability distribution of income $P(r,t)$ obeys the Fokker-Planck equation (\ref{diffusion}) with the variable $m$ is replaced by the variable $r$.  We focus on the stationary distribution (\ref{stationary}), where, again, $m$ is replaced by $r$.  To make further process, we need to make some assumptions about the drift and diffusion coefficients $A(r)$ and $B(r)$.  Empirical analysis of income distribution shows that it has a well-defined two-class structure \cite{Yakovenko-2001b,Silva-2005}.  The lower class is described by the exponential Boltzmann-Gibbs function, which was uncovered in Ref.~\cite{Yakovenko-2001a} and then confirmed for the middle part of income distributions in many countries \cite{Tao-2019}.  In contrast, the upper class is described by the Pareto power law \cite{Pareto-1897}.  The two-class structure of income distribution can be rationalized on the basis of a kinetic approach as follows.

For the lower class, where income comes from wages and salaries, it is reasonable to assume that income changes do not depend on income itself, i.e.,\ $\Delta$ is independent of $r$.  Such a stochastic process is called additive \cite{Silva-2005}.
Then the coefficients in Eq.~(\ref{diffusion}) are some constants $A_0$ and $B_0$ independent of $r$, and the stationary distribution of income (\ref{stationary}) is exponential, analogous to the Boltzmann-Gibbs distribution (\ref{Gibbs})
\begin{equation}  \label{exponential}
  P_{\rm add}(r) = \frac{1}{T}\,e^{-r/T}, \qquad T=\frac{B_0}{A_0}.
\end{equation}
Here the parameter $T$ is the income temperature, which could be denoted as $T_r$, but we drop the subscript $r$ to shorten notation.  It is equal to the mean income of the lower class: $T=\langle r\rangle_{\rm add}$.

In contrast, the upper-class income comes from bonuses, investments, and capital gains, which are calculated as percentages.  Therefore, for the upper class, it is reasonable to expect that income changes are proportional to income itself, i.e.,\
$\Delta\propto r$.  This is known as the proportionality principle of Gibrat \cite{Gibrat-1931}, and such a stochastic process is called multiplicative \cite{Silva-2005}.  Then $A=ar$ and $B=br^2$, and Eq.~(\ref{stationary}) gives a power-law distribution
  \begin{equation}  \label{power-law}
  P_{\rm mult}(r) \propto \frac{1}{r^{1+\alpha}}, \qquad \alpha=1+\frac ab.
  \end{equation}
The multiplicative hypothesis for the upper-class income was quantitatively verified in Ref.~\cite{Aoki-2003} for Japan, where tax identification data are officially published for the top taxpayers.

In practice, the additive and multiplicative processes may coexist.  For example, an employee may receive a cost-of-living raise calculated in percentages (the multiplicative process) and a merit raise calculated in dollars (the additive process).  Assuming that these processes are uncorrelated, we find that $A=A_0+ar$ and $B=B_0+br^2=b(r_0^2+r^2)$, where $r_0^2=B_0/b$.  Substituting these expressions into Eq.~(\ref{stationary}), we find \cite{Yakovenko-2009,Banerjee-2010}
  \begin{equation}  \label{arctan}
  P_{\rm int}(r)=c\,\frac{e^{-(r_0/T)\arctan(r/r_0)}}
  {[1+(r/r_0)^2]^{(1+\alpha_i)/2}} ,
  \end{equation}
The distribution (\ref{arctan}) interpolates between the exponential law for low $r$ and the power law for high $r$, because either the additive or the multiplicative process dominates in the corresponding limit.  A crossover between the two regimes takes place at $r\approx r_0$, where the additive and multiplicative contributions to $B$ are equal.  The distribution (\ref{arctan}) has three parameters: the temperature $T=A_0/B_0$, the Pareto exponent $\alpha_i=1+a/b$, and the crossover income $r_0$.  It is a minimal model that captures the salient features of the two-class income distribution  \cite{Yakovenko-2009,Banerjee-2010}.  The label $i$ distinguishes the exponent $\alpha_i$ in the interpolated formula (\ref{arctan}) from the exponent $\alpha$ in the simple power law (\ref{power-law}) for the purpose of data fitting, as discussed in Sec.~\ref{sec:empirics}\ref{sec:exponent}.  A formula similar to (\ref{arctan}) was also derived by Fiaschi and Marsili \cite{Fiaschi-2012} for a microeconomic model effectively described by Eq.~(\ref{diffusion}).  But Eq.~(\ref{arctan}) was published  for the first time in 1895 in this journal by the statistician Karl Pearson \cite{Pearson-1895} and is known as the Pearson Type IV distribution.

Although Eqs.~(\ref{exponential}), (\ref{power-law}), and (\ref{arctan}) are derived from the stationary solution (\ref{stationary}) of the Fokker-Planck equation, in the next Section we treat them as quasi-stationary and empirically determine how their parameters evolve over time.

\section{Data analysis of income distribution in the USA in 1983--2018}  \label{sec:empirics}

\subsection{The two-class fit of the cumulative distribution function}

\begin{figure}[b]
\centering\includegraphics[width=\linewidth]{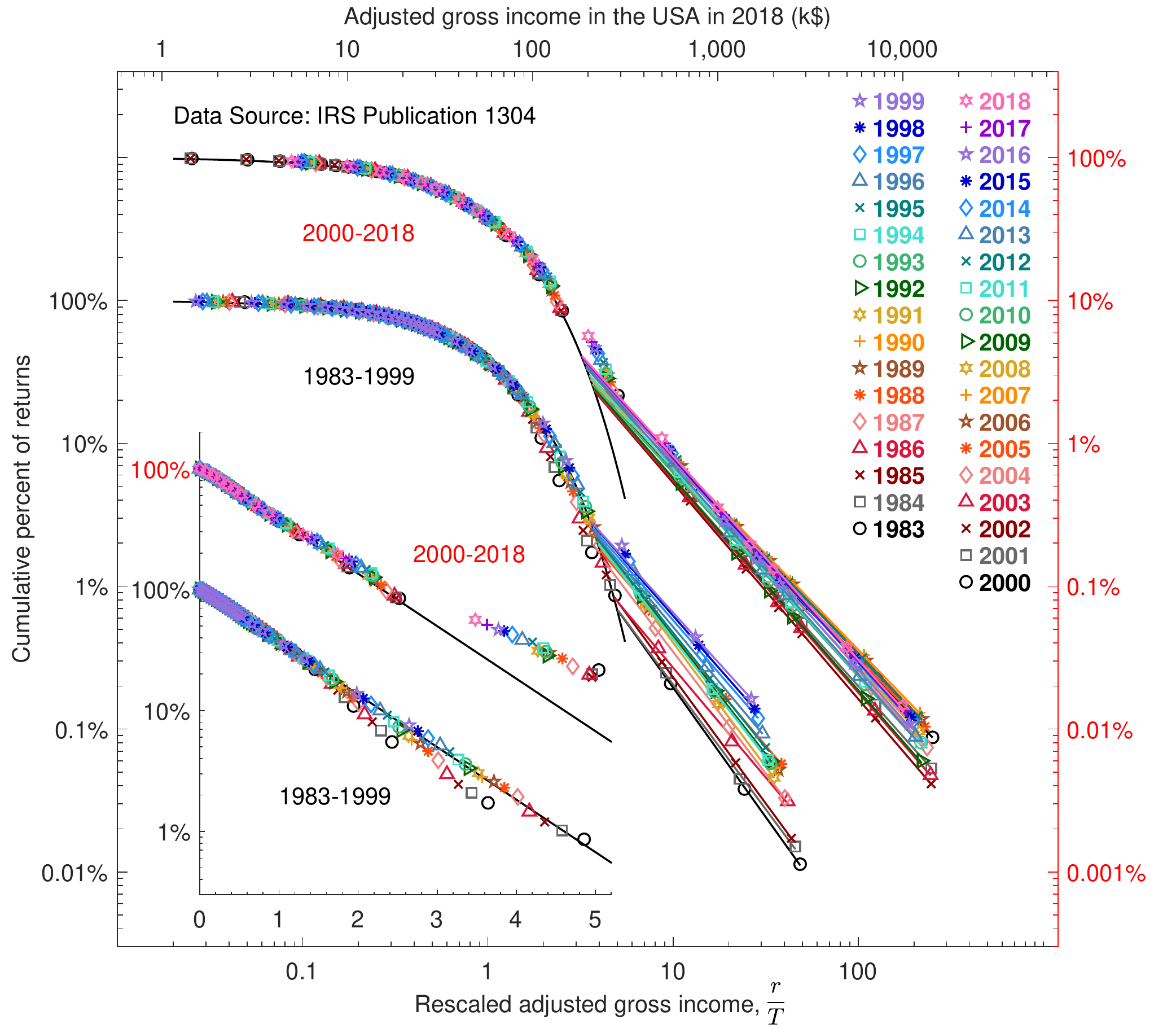}
\caption{The cumulative distribution function $C(r)$ versus the rescaled income $r/T$ in the USA for 1983--2018.  The main panel is in log-log scale, whereas the inset is in log-linear scale.}
\label{Fig:CDF-superimposed}
\end{figure}

We use the variable $r$ for annual income, which is called the adjusted gross income in the IRS Publication 1304 \cite{IRS-1304}.  Income distribution data are reported in this publication at fixed income levels.  At the high end, these levels are \$100k, \$200k, \$500k, and \$1M for the years earlier than 2000.  Here k denotes thousand dollars, and M million dollars.  Starting from 2000, the data for the income levels of \$1.5M, \$2M, \$5M, and \$10M are also reported.  At the low end below \$100k, the data are reported at many income levels.  Further technical details about the source and analysis of the data are provided in Supplemental Material \cite{Supplement}.

Given these fixed income bins set by IRS, we introduce the (complementary) cumulative distribution function $C(r)=\int_r^\infty P(r')\, dr'$ (CDF), obtained by integration of the probability density function $P(r)$ (PDF).  The cumulative function $C(r)$ is the fraction of tax returns reporting adjusted gross income above $r$ in the IRS data.  In our paper, we treat the fraction of tax returns as a proxy for the fraction of population with income above $r$.  This correspondence is only approximate for several reasons.  First, some tax returns are filed jointly by married couples, i.e.,\ one tax return per two people.  The fraction of joint tax returns has dropped from about 1/2 in 1983 to about 1/3 in 2018.  Second, people with income below a certain threshold are not required to file tax returns, but many of them do because of refunds from federal tax withholding and tax credit incentives.  Clearly, the correspondence between the fractions of population and tax returns is not perfect, but it is the best we can do, given the publicly available IRS data.

The cumulative distribution $C(r)$ is given by the exponential and power-law functions for the corresponding probability densities (\ref{exponential}) and (\ref{power-law}):
  \begin{equation}  \label{CDF}
  C_{\rm add}(r) = e^{-r/T},  \qquad \qquad
  C_{\rm mult}(r) \propto \frac{1}{r^\alpha}.
  \end{equation}
We perform a piecewise fit of the empirical data to $C_{\rm add}(r)$ at the low end and to $C_{\rm mult}(r)$ at the high end of income distribution.  The IRS publication reports many data points at the low end, which makes it easier to determine the income temperature $T$ from the fits.  In contrast, only few data points are available at the high end, so the Pareto exponent $\alpha$ is obtained with lower accuracy.

Having determined the temperature $T$, we visualize the fits in Fig.~\ref{Fig:CDF-superimposed} using the rescaled income $r/T$ as the dimensionless coordinate on the horizontal axis.  Then the CDF curves for different years collapse on a single curve at the low end of the distribution.  The inset in Fig.~\ref{Fig:CDF-superimposed} shows the data in the log-linear scale, with the rescaled income $r/T$ in linear scale on the horizontal axis and CDF in logarithmic scale on the vertical axis.  The collapse of the data points on a straight line in the log-linear scale indicates that $C(r)$ is, indeed, well described by the exponential function.  For clarity, the data for 1983--1999 and 2000--2018 are separated in the inset.  The data points for the early 1980s deviate slightly below the exponential line, indicating lower inequality in those years, but this deviation subsequently goes away.  The upper boundary of the exponential fit gradually shrinks from $r/T\approx3.5$ in the late 1980s to $r/T\approx2.5$ by 2000, with data points at the higher incomes deviating upwards for 2000--2018 in the inset.

The main panel in Fig.~\ref{Fig:CDF-superimposed} shows the same data in log-log scale, where both axes for $r/T$ and CDF are logarithmic.  At the lower end, the data points collapse on the black curve representing the exponential distribution.  At the high end, the data points fall on straight lines, representing the power law in log-log coordinates.  The slopes of these lines give the Pareto exponent $\alpha$.  We observe a noticeable decrease in the Pareto slope from 1983 to 1999, which signals fattening of the upper tail, but then a narrower variation of the slope in 2000--2018.  While the main panel in Fig.~\ref{Fig:CDF-superimposed} presents clear data-based evidence for the two-class society, it also demonstrates the absence of the ``middle class''.  There is no objective, commonly agreed upon definition of the ``middle class'' concept in the economic literature, where different authors make up their own subjective definitions of the ``middle class''.

\begin{figure}[b]
\centering\includegraphics[width=0.55\linewidth]{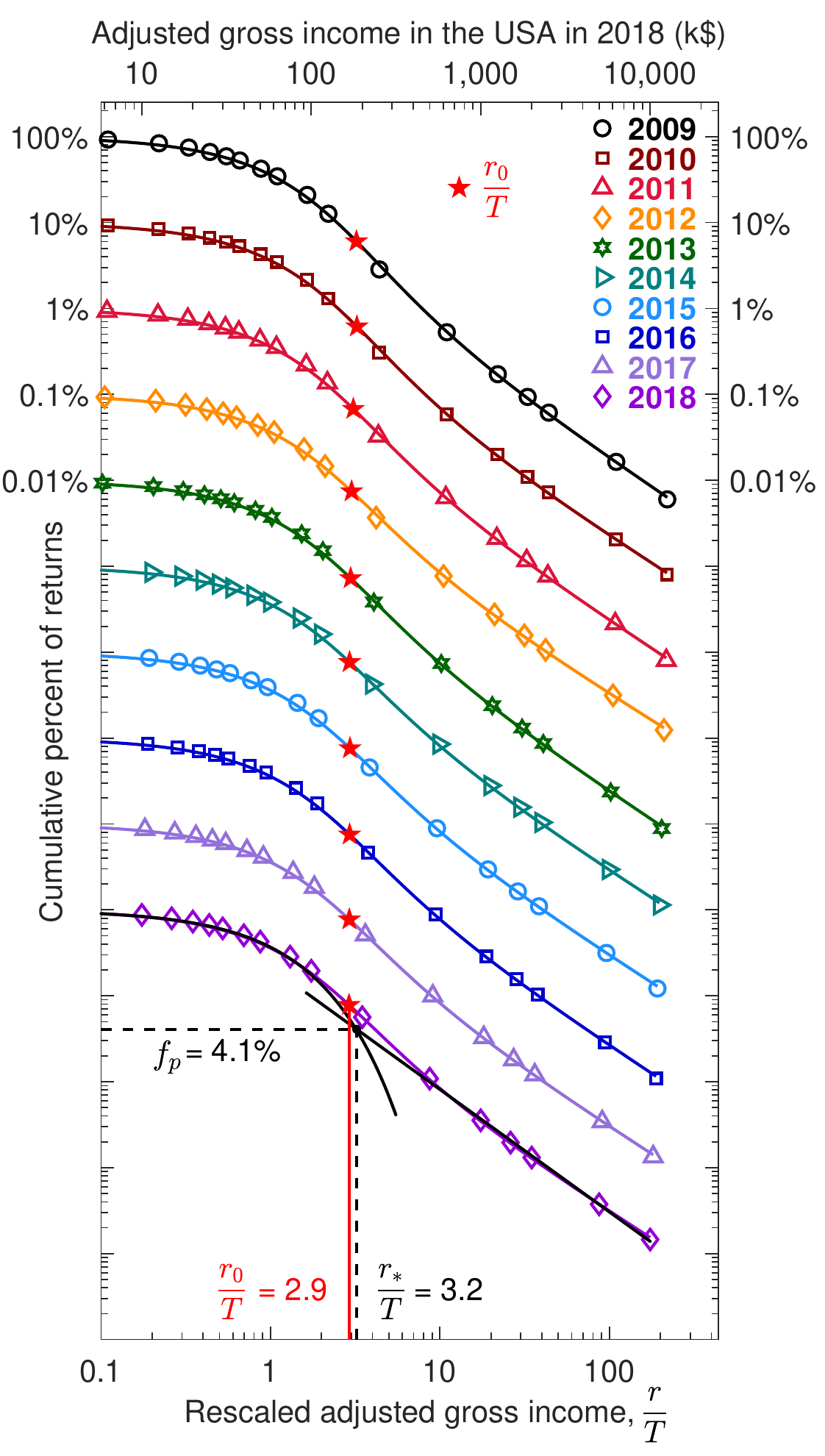}
\caption{The cumulative distribution function $C(r)$ versus the rescaled income $r/T$ in the USA for 2009--2018 fitted to the interpolating function (\ref{arctan}).  The crossover income $r_0$ of the interpolating fit is indicated by the red star.  The small black circle indicates another crossover parameter $r_*$ obtained by piecewise fit.}
\label{Fig:CDF-arrayed}
\end{figure}

As an alternative to the piecewise fit, in Fig.~\ref{Fig:CDF-arrayed} we show fits of the CDF data for the last ten years 2009--2018 to the continuous interpolated function (\ref{arctan}) in log-log scale.  We first determine $T$ by fitting the low-end data as described above.  Then we determine the Pareto exponent $\alpha_i$ and the crossover income level $r_0$ by fitting to Eq.~(\ref{arctan}).  Because of the different fitting procedure, the value of the Pareto exponent $\alpha_i$ obtained from the interpolated fit is slightly different from the value $\alpha$ obtained from the piecewise fit.

The crossover point $r_0$ between the exponential and power-law distributions is indicated in Fig.~\ref{Fig:CDF-arrayed} by the red star.  An alternative crossover point $r_*$ is obtained as an intersection (indicated by the small black circle) of the piecewise exponential and power-law fits shown by black curves for 2018.  The values of $r_0$ and $r_*$ are generally different, but rather close.  We perform the piecewise fit and determine $r_*$ for all available years.  In contrast, the interpolated fit is performed only for the last ten years 2009--2018, where it works well.  For the earlier years, it does not work well, because the crossover between the exponential and power-law distributions is less smooth and more cuspy, as seen in Fig.~\ref{Fig:CDF-superimposed}.  Moreover, the lower part deviates downward from the exponential fit in the early 1980s.

\begin{table}[t]  
\caption{Parameters of the fit for 2018.  Notation is explained in the text.}
\label{Tab:parameters}
\begin{tabular}{ccccccccccccc}
\hline
$r_{\rm med}$ & $T$ & $\langle r\rangle$ & $r_0$ & $r_*$ & $\alpha$ & $\alpha_i$ & $f_p$ & $f_r$ & $f_t$ & $f_L$ & G & G$_{\rm tax}$\\
\$41k & \$57k & \$78k & \$167k & \$185k & 1.42 & 1.25 & 4\% & 34\% & 58\% & 22\% & 60\% & 82\% \\
\hline
\end{tabular}
\vspace*{-4pt}
\end{table}

The two-class decomposition allows us to characterization the empirical income distribution by a small set of fitting parameters.  Their values are shown in Table \ref{Tab:parameters}
for the latest year 2018 and discussed in subsequent sections, where we visualize the historical evolution of these parameters.

\subsection{Divergence of the mean and the median incomes}

\begin{figure}[b]
\centering\includegraphics[width=0.7\linewidth]{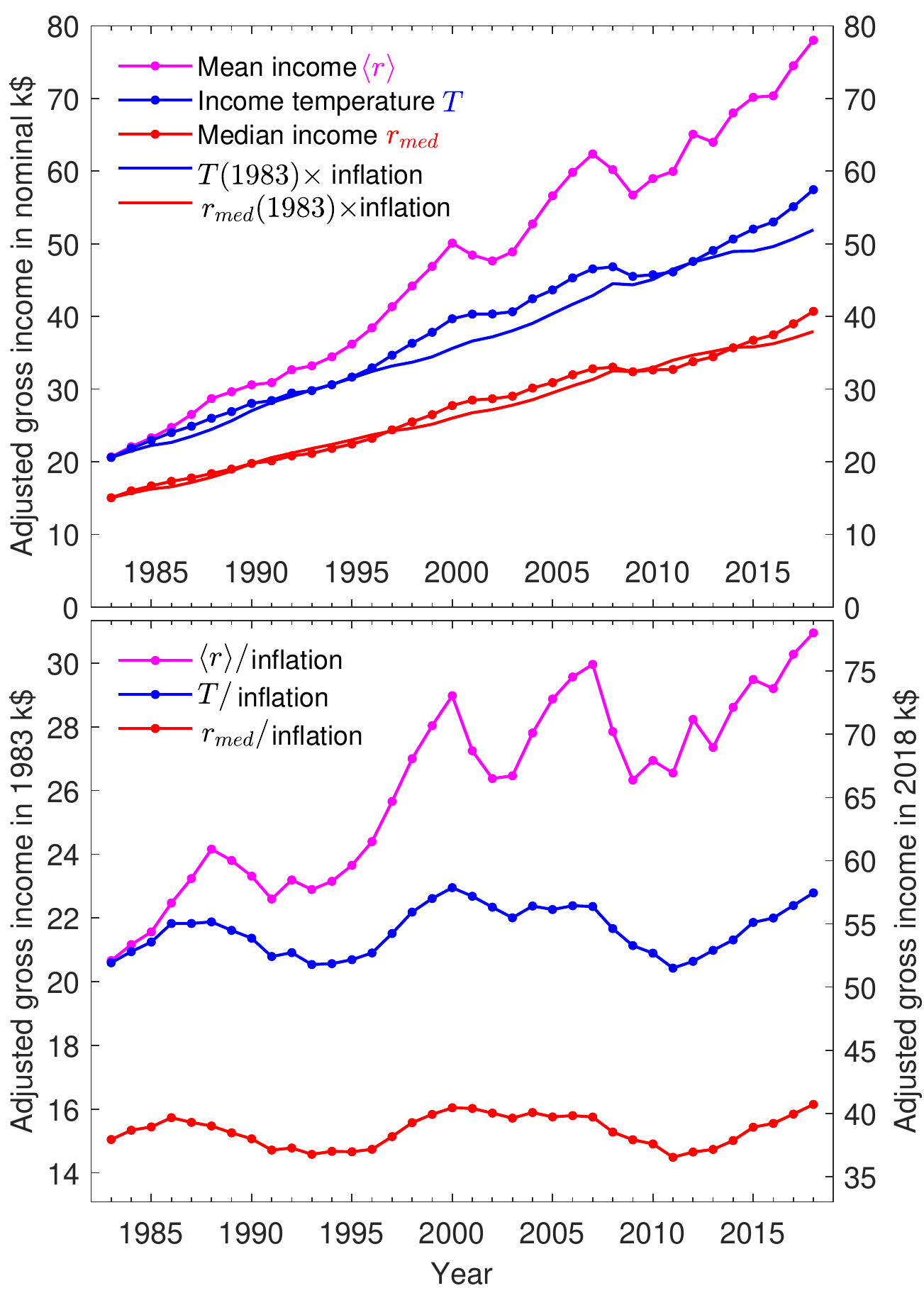}
\caption{The median $r_{\rm med}$ and mean $\langle r\rangle$ incomes and the income temperature $T$, which is the mean income of the lower class, in nominal dollars (top panel) and divided by inflation (bottom panel).}
\label{Fig:mean-median-T}
\end{figure}

The dotted curves in the top panel of Fig.~\ref{Fig:mean-median-T} shows the historical evolution of the income temperature $T$, as well as the mean $\langle r\rangle=\int_0^\infty dr\,r\,P(r)$ and median $r_{\rm med}$ incomes.  All of them increase in time in nominal dollars, but some of that increase is due to inflation.  We define the inflation parameter as the ratio CPI(year)/CPI(1983) of the Consumer Price Index (CPI) for a given year to the CPI in 1983 \cite{CPI}.  The solid curves in the top panel of Fig.~\ref{Fig:mean-median-T} show the parameters $T$ and $r_{\rm med}$ taken in 1983 and multiplied by the inflation parameter for the subsequent years.  The solid curves are close to the dotted curves for $T$ and $r_{\rm med}$, so the increase in nominal dollars of the income temperature $T$ and the median income $r_{\rm med}$ is primarily due to inflation.

The bottom panel of Fig.~\ref{Fig:mean-median-T} shows the parameters $\langle r\rangle$, $T$, and $r_{\rm med}$ divided by the inflation parameter.  Again, we observe that the parameters $T$ and $r_{\rm med}$ characterizing the lower class remain approximately constant over 36 years, when adjusted for inflation.  For the exponential distribution in Eq.~(\ref{CDF}), these two parameters are simply proportional: $r_{\rm med}=T\ln2 \approx 0.7\,T$.

In contrast, the overall mean income $\langle r\rangle$ increases sharply relative to inflation.  The divergence between the income temperature $T=\langle r\rangle_{\rm add}$, which is the mean income of the lower class, and overall mean income $\langle r\rangle$, which includes the upper class, points to a significant increase of inequality in the last 36 years.  As the bottom panel of Fig.~\ref{Fig:mean-median-T} shows, these two mean incomes were approximately equal in 1983, but have diverged since then.  Thus, the growth of inequality is primarily driven the increase of the upper-class income relative to the lower class, while the lower-class income remains approximately constant when adjusted for inflation.

\begin{figure}[b]
\centering\includegraphics[width=0.8\linewidth]{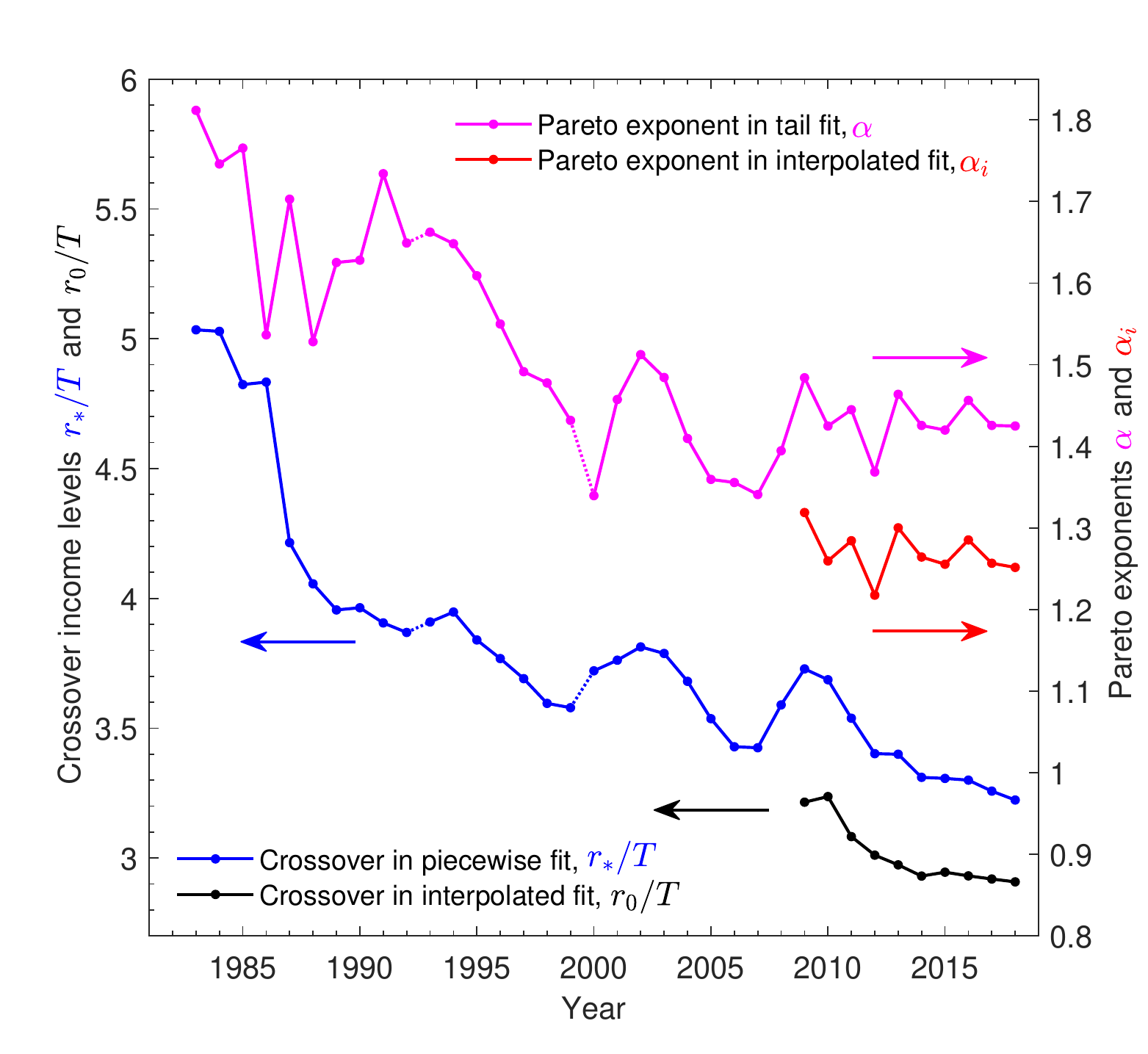}
\caption{Top two curves: the Pareto exponents $\alpha$ from the piecewise fits and $\alpha_i$ from the interpolated fits.  Bottom two curves: the crossover incomes $r_*$ (from the piecewise fits) and $r_0$ (from the interpolated fits) separating the lower and upper classes, normalized to the income temperature $T$.}
\label{Fig:exponent-crossover}
\end{figure}

\subsection{The Pareto exponent and the boundary between two classes}
\label{sec:exponent}

The top two curves in Fig.~\ref{Fig:exponent-crossover} show the historical evolution of the Pareto exponents $\alpha$, obtained by the piecewise fits for all years, and $\alpha_i$, obtained by the interpolated fits for the last 10 years using Eq.~(\ref{arctan}).  The values of $\alpha$ and $\alpha_i$ are slightly different because they are obtained by different fitting procedures, but they exhibit similar behavior.  We observe a significant decrease of the exponent $\alpha$ from 1983 to 2000, which indicates fattening of the tail.  But after 2000, the exponent $\alpha$ exhibits a ``bumpy plateau'', where it decreases during financial bubbles and increases during crashes, but stays approximately constant overall.  These observations are qualitatively consistent with the main panel in Fig.~\ref{Fig:CDF-superimposed}, as well as Fig.~\ref{Fig:mean-median-T}, and indicate that the sharp expansion of the upper class happened in the 1980s and 1990s, but have slowed down or saturated after 2000.

The two curves at the bottom of Fig.~\ref{Fig:exponent-crossover} show the historical evolution of the crossover incomes $r_*$ and $r_0$ separating the lower and upper classes, as illustrated in Fig.~\ref{Fig:CDF-arrayed}.  The former is obtained from the piecewise fits for all years, while the latter from the interpolated fits for the last 10 years, and both are normalized to the income temperature $T$.  The ratio $r_*/T$ has decreased from about 5 in 1983 to slightly above 3 in 2018.  A particularly sharp drop is visible around 1986, coinciding with the Tax Reform Act of the Reagan administration.  The decreasing value of the ratio $r_*/T$ indicates that the exponential part of the distribution is shrinking, whereas the power-law part is expanding, which is consistent with the inset in Fig.~\ref{Fig:CDF-superimposed}.

\subsection{The income, population, and tax fractions of the upper class}

\begin{figure}[b]
\centering\includegraphics[width=0.7\linewidth]{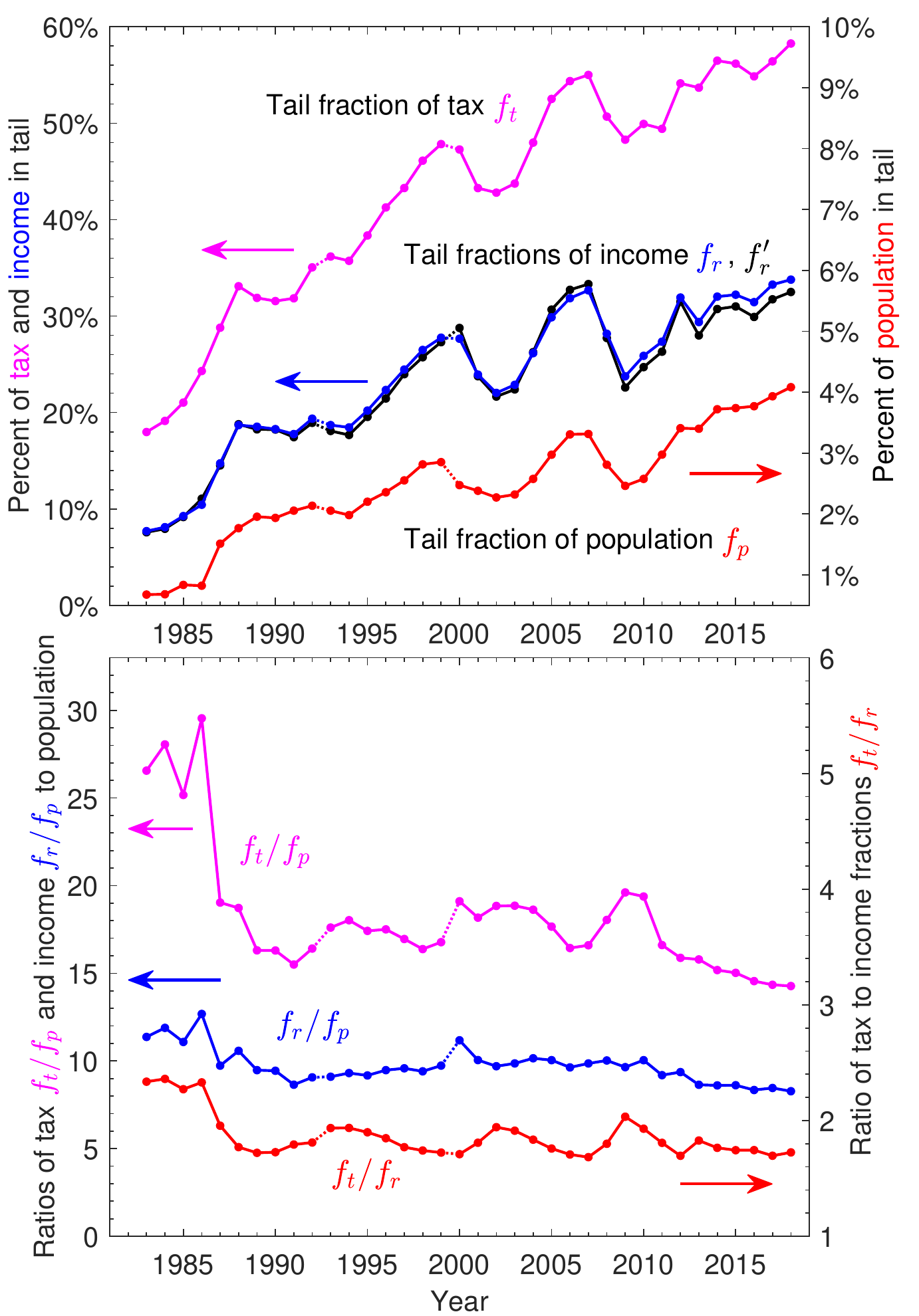}
\caption{Top panel: the fractions of upper-class population $f_p$, income $f_r$, and tax paid $f_t$ relative to the total population.  The income fraction $f_r$ is deduced directly from the IRS data, whereas $f_r'$ is calculated from Eq.~(\ref{upper-share}).  Bottom panel:  the ratios of these fractions.}
\label{Fig:fractions}
\end{figure}

Given the boundary $r_*$ between the two classes, we can also characterize the upper class by the fractions of its population $f_p$, income $f_r$, and tax paid $f_t$ relative to the total population:
  \begin{equation}  \label{fractions}
  f_p=\int_{r_*}^\infty dr\,P(r), \qquad 
  f_r=\frac{1}{\langle r\rangle} \, \int_{r_*}^\infty dr\,r\,P(r),
  \end{equation}
whereas $f_t$ is defined similarly to $f_r$ but for the tax paid rather than income.  The historical evolution of these fractions is shown in the top panel of Fig.~\ref{Fig:fractions}.  All of these fractions increase in time, indicating expansion of the upper class.  In particular, the lower curve in the top panel shows that the upper-class population share $f_p$ has increased from less than 1\% in 1983 (when almost the whole income distribution was close to exponential) to slightly greater than 4\% in 2018.  The increase in the upper-class population fraction is closely related to the decrease in the crossover income ratio $r_*/T$ separating the two classes, as shown by the bottom curves in Fig.~\ref{Fig:exponent-crossover}.

The income fraction $f_r$ of the upper class exhibits the ``bumpy plateau'' pattern in the top panel of Fig.~\ref{Fig:fractions} for the last 20 years, similarly to the exponent $\alpha$ in Fig.~\ref{Fig:exponent-crossover} and the mean income $\langle r\rangle$ in Fig.~\ref{Fig:mean-median-T}.  We observe sharp peaks of the upper-class income share $f_r$ around 2000 and 2017 coinciding with the stock market and real estate bubbles, followed by crashes of these bubbles, and then further increase by 2012, under the quantitive easing policy of the Federal Reserve Bank.  Clearly, the upper-class income share is closely related to financial bubbles. 

The income fraction $f_r$ shown in Fig.~\ref{Fig:fractions} is deduced directly from the IRS data, given the income threshold $r_*$.  As a consistency check, we compare it with the theoretically expected fraction $f_r'$ for the power-law distribution (\ref{power-law}).  The expected mean income of the upper class and the income fraction relative to the total population are
  \begin{equation}  \label{upper-share}
  \langle r\rangle_{\rm mult} 
  = \frac{\int_{r_*}^\infty dr\,r/r^{1+\alpha}}{\int_{r_*}^\infty dr/r^{1+\alpha}}
  = \frac{\alpha}{\alpha-1}\,r_*, \qquad \qquad
  f_r' = f_p \, \frac{\langle r\rangle_{\rm mult}}{\langle r\rangle}
  = f_p \, \frac{\alpha}{\alpha-1}\,\frac{r_*}{\langle r\rangle}.
  \end{equation}
The upper-class income nominally diverges, $f_r'\to\infty$, when $\alpha\to1$.  When extrapolated from the decreasing trend in 1980s and 1990s in Fig.~\ref{Fig:exponent-crossover}, the exponent $\alpha$ could have reached 1 by 2018, but saturated after 2000 at the ``bumpy plateau'' of bubbles and crashed, thus avoiding the singularity.

The expected income fraction $f_r'$ calculated from Eq.~(\ref{upper-share}) with the parameters from the piecewise fits is shown in the top panel of Fig.~\ref{Fig:fractions} together with $f_r$.  We see that $f_r$ and $f_r'$ are very close for most years, indicating that the power law (\ref{power-law}) accounts for all of the upper-class income.  For the recent years, we notice that $f_r>f_r'$, suggesting that the actual upper-class income may be slightly higher relative to the power-law extrapolation.  The difference between $f_r$ and $f_r'$ could indicate the additional income of oligarchy on top of the power law, as proposed for the wealth distribution in Ref.~\cite{Boghosian-2017}.  However, the difference in Fig.~\ref{Fig:fractions} is too small to be taken as evidence for oligarchy in income distribution, given the uncertainty of the fitting parameters.

The ratios of the fractions are shown in the bottom panel of Fig.~\ref{Fig:fractions}.  The middle curve shows the ratio $f_r/f_p$ of the income and population fractions.  This ratio stays approximately constant $\approx 10$ and even slightly decreases over time.  Comparison with Eq.~(\ref{upper-share}) for $f_r'/f_p$ suggests that the decrease in $r_*/\langle r\rangle$ is compensated by the increase in $\alpha/(\alpha-1)$, as shown in Fig.~\ref{Fig:exponent-crossover}.

Another indicator is the share of taxes $f_t$ paid by the upper class, relative to its income share $f_r$.  The lower curve in the bottom panel of Fig.~\ref{Fig:fractions} shows a noticeable decrease in $f_t/f_r$ after the tax reform of 1986, but then it stays approximately constant $\approx 2$.  Relatedly, the top panel shows that, after 2000, the income share $f_r$ oscillates roughly between 20\% and 30\%, whereas the tax share $f_t$ between 40\% and 60\%.  Notably, the upper class paid close to 60\% of taxes in 2018.  The top curve in the bottom panel shows the ratio $f_t/f_p$ of tax to population fractions for the upper class.  It exhibits a sharp drop around 1986 and then a gradual decrease in the last 10 years.

\subsection{Lorenz curves and Gini coefficient}

\begin{figure}[b]
\centering\includegraphics[width=0.8\linewidth]{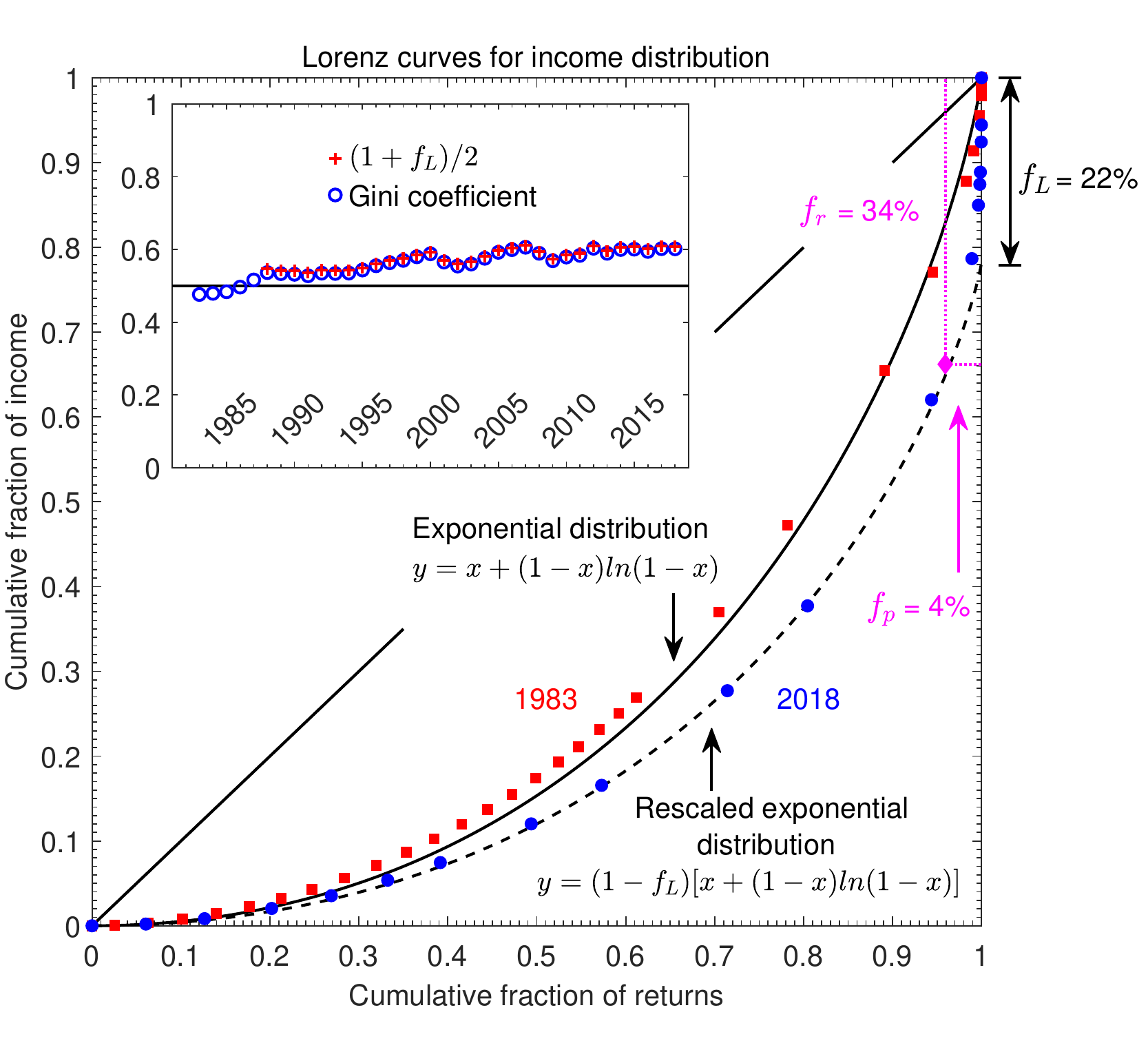}
\caption{The IRS data points for the Lorenz curves of income distribution in 1983 (red squares) and 2018 (blue circles).  The solid and dashed curves represent the exponential, Eq.~(\ref{Lorenz_exp}), and rescaled exponential, Eq.~(\ref{Lorenz_jump}), distributions characterizing the lower class.  The population $f_p$ and income $f_r$ fractions mark the boundary of the upper class in 2018.  The inset shows the empirical Gini coefficient in comparison with Eq.~(\ref{Gini-f}).}
\label{Fig:Lorenz-income}
\end{figure}

Income distribution is commonly characterized by the Lorenz curve in the economic literature \cite{Kakwani-1980}.  The Lorenz curve is defined parametrically in terms of the two coordinates $x(r)$ and $y(r)$
  \begin{equation}
  x(r) = \int_0^r dr'P(r'), \quad
  y(r) = \frac{1}{\langle r\rangle} \, \int_0^r dr'r'P(r').
  \label{Lorenz}
  \end{equation}
Here $x(r)$ is the fraction of the population with incomes below $r$, and $y(r)$ is the total income of this population, as a fraction of the total income in the system.  When the parameter $r$ changes from 0 to $\infty$, the variables $x(r)$ and $y(r)$ change from 0 to 1, thus producing the Lorenz graph in the $(x,y)$ plane.  Its advantage is that all available data are represented within a finite area in the $(x,y)$ plane, whereas the upper tail at $r\to\infty$ is inevitably truncated in Figs.~\ref{Fig:CDF-superimposed} and \ref{Fig:CDF-arrayed}.

It was derived in Ref.~\cite{Yakovenko-2001a} that the Lorenz curve for the exponential distribution (\ref{exponential}) is 
  \begin{equation}  \label{Lorenz_exp}
  y=x+(1-x)\ln(1-x).
  \end{equation}
When the upper class is present, the Lorenz curve for the lower class is obtained \cite{Silva-2005} by multiplying Eq.~(\ref{Lorenz_exp}) by the factor $(1-f_L)<1$:\begin{equation}  \label{Lorenz_jump}
  y = (1-f_L)[x + (1-x) \ln(1 -x)].
  \end{equation}
The factor $(1-f_L)$ is the fraction of total income contained in the extrapolated exponential distribution, whereas $f_L$ is the additional fraction of income in the upper class.  (The subscript $L$ alludes to the Lorenz curve.)

The red squares and blue circles in Fig.~\ref{Fig:Lorenz-income} show the IRS data points for the Lorenz curves of income distribution in 1983 and 2018.  The straight diagonal line represents the reference case of perfect equality, so inequality is higher when the data points are further away from it.  We observe that the Lorenz curve for 1983 is very close to the solid black curve representing Eq.~(\ref{Lorenz_exp}) for the exponential distribution.  Moreover, the data points in the middle are slightly above the solid black curve, indicating lower inequality than in the exponential distribution.  These conclusions are consistent with the CDF graphs in Fig.~\ref{Fig:CDF-superimposed} and the observation that $\langle r\rangle\approx T$ for 1983 in Fig.~\ref{Fig:mean-median-T}.

In contrast, the Lorenz curve for 2018 is shifted down, signaling an increase in inequality.  Most of its data points, except at the high end, are well fitted by the rescaled exponential formula (\ref{Lorenz_jump}) shown by the dashed black curve.  This indicates that income distribution within the lower class remains exponential, whereas the increase in inequality comes from the excess share $f_L$ of the upper-class income, determined by the intersection of the fit (\ref{Lorenz_jump}) with the vertical axis.  

\begin{figure}[b]
\centering\includegraphics[width=0.8\linewidth]{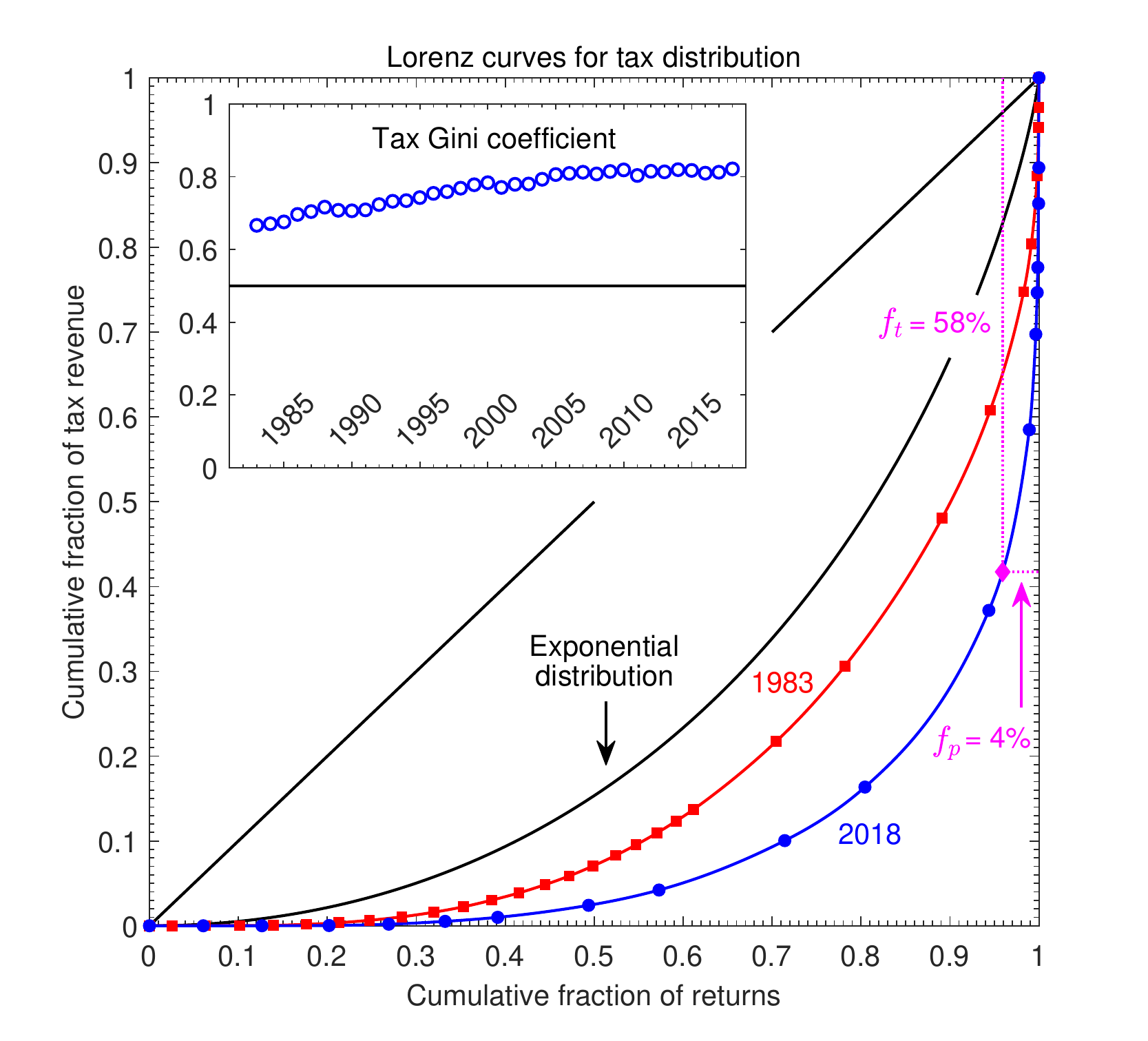}
\caption{The Lorenz curves of tax distribution in 1983 (red squares) and 2018 (blue circles).  The black solid curve represents the exponential distributions, Eq.~(\ref{Lorenz_exp}), for a reference.  The population $f_p$ and tax $f_t$ fractions of the upper class are indicated for 2018.  The inset shows the tax Gini coefficient.}
\label{Fig:Lorenz-tax}
\end{figure}

Also shown in Fig.~\ref{Fig:Lorenz-income} are the population $f_p$ and income $f_r$ fractions of the upper class.  Both parameters $f_L$ and $f_r$ represent an income share of the upper class, but they are defined differently.  In Fig.~\ref{Fig:Lorenz-income}, the parameter $f_r=34\%$ is the income share of the top 4\% of population, whose income distribution follows a power law.  In contrast, $f_L=22\%$ is the extra income of the upper class relative the exponential distribution extrapolated to $r\to\infty$.  Thus, the parameter $f_L$ measures a deviation of the actual income distribution from the purely exponential one.

To characterize inequality by a single number, the economic literature \cite{Kakwani-1980} uses the Gini coefficient $0\leq G\leq1$, defined as twice the area between the diagonal line and the Lorenz curve.  It was shown that $G=1/2$ for the exponential distribution \cite{Yakovenko-2001a}, and 
  \begin{equation}  \label{Gini-f}
  G=\frac{1+f_L}{2}
  \end{equation}
when taking into account the fraction $f_L$ of the upper-class income on top of the exponential distribution \cite{Silva-2005}.  The values of $G$ deduced from the IRS data are shown by the open circles in the inset of Fig.~\ref{Fig:Lorenz-income}, where the crosses show the values calculated by Eq.~(\ref{Gini-f}) starting from 1988.  The agreement between the empirical values of $G$ and the formula (\ref{Gini-f}) in Fig.~\ref{Fig:Lorenz-income} demonstrates that the increase in income inequality starting from the late 1980s comes completely from the growth of the upper income share relative to the lower class, whereas income inequality within the lower class itself remains constant and exponentially distributed.

Besides the overall increase since 1983, the inset in Fig.~\ref{Fig:Lorenz-income} exhibits local maxima corresponding to bubbles in financial markets.  The Gini coefficient $G$ peaks around 1988 during the bubble in Savings and Loan, around 2000 in the stock market, around 2007 in subprime mortgages, and past 2012 under quantitative easing.  Thus we confirm that the overall income inequality, as measured by the Gini coefficient, peaks during bubbles in financial markets.  This behavior corresponds to the ``bumpy plateau'' already discussed for Figs.~\ref{Fig:mean-median-T}, \ref{Fig:exponent-crossover}, and \ref{Fig:fractions}.

\begin{figure}[b]
\centering\includegraphics[width=0.8\linewidth]{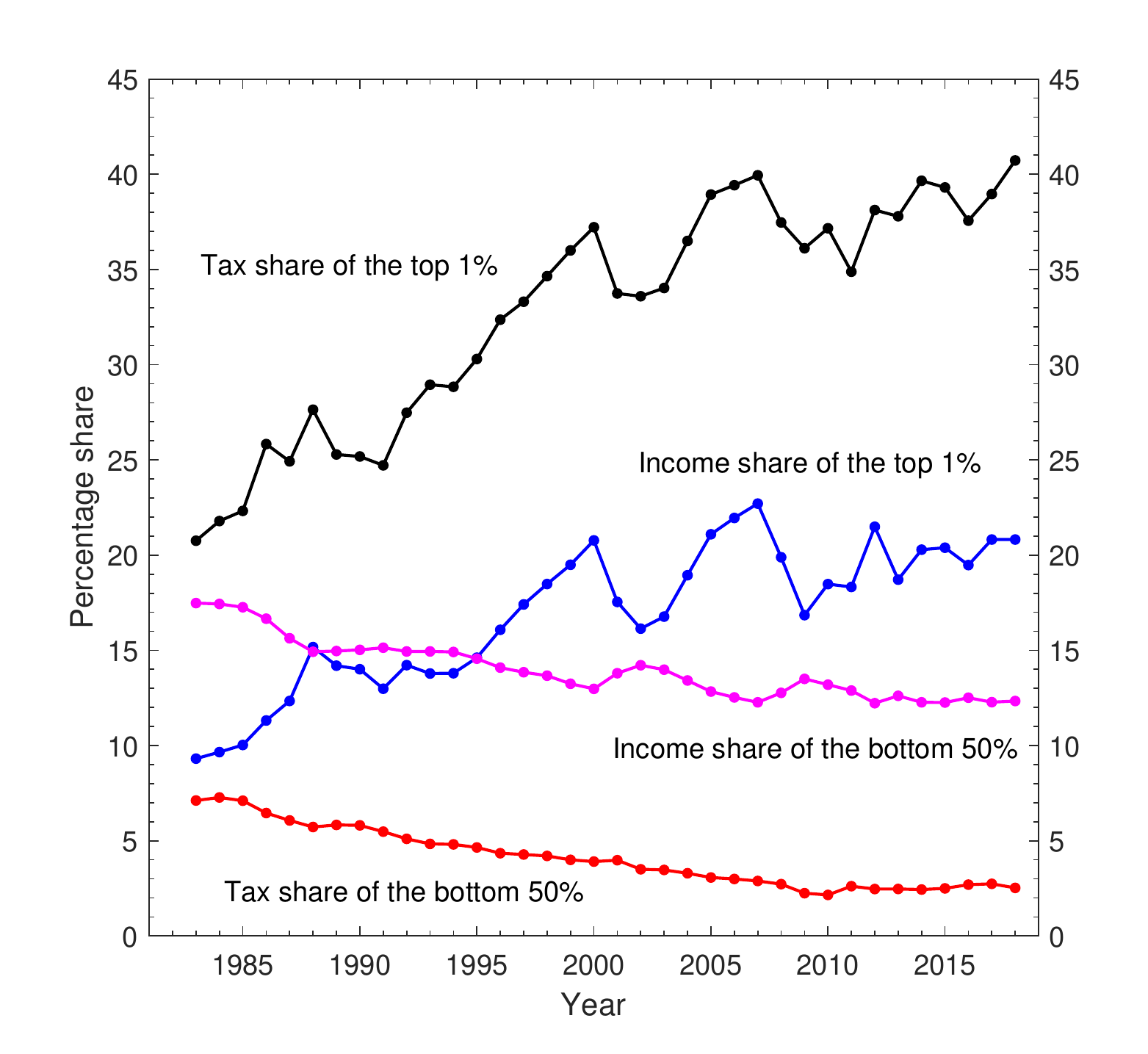}
\caption{Income and tax shares of the bottom 50\% and the top 1\% of population.}
\label{Fig:50-1-percent}
\end{figure}

The IRS publication \cite{IRS-1304} also provides information on the distribution of taxes paid by the population.  In Fig.~\ref{Fig:Lorenz-tax} we show the Lorenz curves for tax revenue in 1983 and 2018, along with the Gini coefficient for taxes in the inset.  Because of progressive taxation, where the recipients of higher income are subject to higher tax rates, the inequality of tax distribution in Fig.~\ref{Fig:Lorenz-tax} is higher than in Fig.~\ref{Fig:Lorenz-income} for income distribution, as also exemplified by the higher tax Gini coefficient.  We observe that the tax revenue became highly concentrated in the upper class by 2018, because of the relative decrease in the lower-class income due to the overall increase in income inequality.  As shown in Fig.~\ref{Fig:Lorenz-tax}, the upper class, consisting of only about $f_p=4\%$ of population, contributed the fraction $f_t=58\%$, greater than a half, of the total tax revenue, as also shown by the top curve in Fig.~\ref{Fig:fractions}.

To amplify these observations, we show the income and tax shares of the bottom 50\% and the top 1\% of population in Fig.~\ref{Fig:50-1-percent}, as derived from the Lorenz curves for income and taxes.  From 1983 to 2018, the income share of the lower half of population has decreased from about 18\% to 12\%, whereas the income share of the top 1\% of population has doubled from about 9\% to 21\%.  At the same time, the sum of their income shares remained approximately constant, increasing only slightly from 27\% in 1983 to 33\% in 2018.  For this reason, the two middle curves in Fig.~\ref{Fig:50-1-percent} look somewhat like mirror images of each other.  Thus, the income share increase of the top 1\% of population mostly came from the income share decrease of the bottom 50\% of population.

During the same years, the tax share of the bottom 50\% of population has decreased from about 7\% to 3\%.  In 2018, the total tax contribution of the lower half of the population is almost negligible, because their total income share is so low.  In contrast, , the tax share of the top 1\% of population has doubled from about 20\% to 40\%.  Thus, the majority of tax revenue now comes from the top few percent of population, where most of income is concentrated.  

A proposal for increasing taxes on the top 1\% of population and incomes above approximately \$500k is currently debated by the Democratic Party of the United States \cite{WashPost-2021}.  The quantitative characterization of the two-class structure of income distribution presented in our paper could provide useful input for the formulation of a tax policy.  In particular, it would be reasonable to align tax brackets with the boundary between the two classes.

\section{Conclusions}  \label{sec:conclusions}

In this paper, income distribution data for the USA in 1983--2018 is analyzed in terms of a two-class decomposition, where the lower class is described by an exponential function, and the upper class by a power law, as inspired by statistical physics.  The quantitative characterization of the two classes presented here may be useful for the formulation of a tax policy.  We observe a sharp increase of income inequality from the early 1980s to the late 1990s, but slower increase or saturation after 2000.  The inequality pattern after 2000 exhibits a ``bumpy plateau'' pattern, with ups and downs related to financial bubbles and crashes, but an approximately constant overall level.

The two-class analysis shows that the growth of inequality originates primarily from the increasing share of income ($f_r$ or $f_L$) in the upper class, relative to the lower class.  At the same time, relative inequality within the lower class, encompassing more than 90\% of population, remains constant and exponentially distributed.  Moreover, the mean income $T$ of the lower class and the median income $r_{\rm med}$ remain approximately constant over 36 years when adjusted for inflation, whereas the overall mean income $\langle r\rangle$ increases relative to inflation.  The growing divergence between the mean and median incomes is another manifestation of increasing inequality.

Not only does the upper-class income share $f_r$ increase, but the fraction $f_p$ of population belonging to the upper class increases too.  In a relative sense, the upper-class population expands (while still remaining a small fraction of only about 4\% of the total population), while the lower-class population shrinks (while still remaining the overwhelming majority).  This happens because the ratio $r_*/T$ of the crossover income $r_*$ separating the two classes to the mean income of the lower class $T$ has significantly decreased from 1983 to 2018.  Thus, the relative income level threshold for getting into the upper class has decreased.

These empirical observations prompt a descriptive interpretation.  We speculate that the significant increase in the upper-class population fraction $f_p$, as well as their income fraction $f_r$, is due to digitization of the economy in the last 40 years.  There was a rapid proliferation of personal computers during the 1980s, followed by the spread of the Internet and the World Wide Web in the 1990s, and then by ubiquitous personal mobile devices and a shift to cloud computing at the present time.  This transformation enabled the creation and relatively easy scaling-up of digital platforms for highly non-local business operations.  In the past, many business were local.  For example, each town had a few local taxi companies, several local book stores, video rental stores, etc.  Now they are largely displaced by a small number of national and global network platforms, such as Uber and Lyft for riding, Amazon for books initially and then for all kinds of goods, Netflix for DVD rentals and video streaming, etc.  The founders and owners of such network platforms become super-rich, because these platforms serve a huge number of customers, in contrast to the old-fashioned local businesses.  Thus we speculate that the growth of the upper class is due to the ongoing transformation of business network topology from local clusters to highly connected superclusters and global hubs.  In contrast to the one-to-one interaction exemplified by Eq.~(\ref{m1+m2}), networked businesses are characterized by the one-to-many interaction topology, resulting in the power law distribution for the upper class.

\vskip6pt

\enlargethispage{20pt}


\dataccess{Supporting data are accessible in \cite{Supplement}.}

\aucontribute{DL performed data analysis and produced the data-based figures under the guidance of VY, then VY wrote the manuscript.  All authors proofread and approved the manuscript.}

\competing{The authors declare that they have no competing interests.}






\begin{thebibliography}{99}

\bibitem{Stanley-1996} H. E. Stanley et al., ``Anomalous fluctuations in the dynamics of complex systems: from DNA and physiology to econophysics'', Physica A \textbf{224} (1996) 302, \url{https://doi.org/10.1016/0378-4371(95)00409-2}

\bibitem{Chakrabarti-history} B. K. Chakrabarti (2005) ``Econophys-Kolkata: a short story'', in \textit{Econophysics of Wealth Distributions}, edited by A. Chatterjee, S. Yarlagadda, and B. K. Chakrabarti (Springer, Milan, 2005), pp 225--228

\bibitem{Redner-1998} S. Ispolatov, P. L. Krapivsky, and S. Redner ``Wealth distributions in asset exchange models'', Eur. Phys. J. B \textbf{2} (1998) 267, \url{https://doi.org/10.1007/s100510050249}

\bibitem{Yakovenko-2000} A. A. Dr\u{a}gulescu and V. M. Yakovenko, ``Statistical mechanics of money'', Eur. Phys. J. B \textbf{17} (2000) 723, \url{https://doi.org/10.1007/s100510070114}

\bibitem{Chakraborti-2000} A. Chakraborti and B. K. Chakrabarti, ``Statistical mechanics of money: how saving propensity affects its distribution'',  Eur. Phys. J. B \textbf{17} (2000) 167, \url{https://doi.org/10.1007/s100510070173}

\bibitem{Bouchaud-2000} Bouchaud JP, M\'ezard M, ``Wealth
condensation in a simple model of economy'', Physica A \textbf{282} (2000) 536, \url{https://doi.org/10.1016/S0378-4371(00)00205-3}

\bibitem{Yakovenko-2009} V. M. Yakovenko and J. B. Rosser Jr., ``Colloquium: Statistical mechanics of money, wealth, and income,'' Rev. Mod. Phys. \textbf{81} (2009) 1703, \url{https://doi.org/10.1103/RevModPhys.81.1703}

\bibitem{Cockshott-2009} Cockshott, W. P.,  A. F. Cottrell, G. J. Michaelson, I. P. Wright, and V. M. Yakovenko, 2009, \textit{Classical Econophysics} (Routledge, Oxford) ISBN 978-0-203-87754-8

\bibitem{Chakrabarti-2013} B. K. Chakrabarti, A. Chakraborti, S. R. Chakravarty, and A. Chatterjee, \textit{Econophysics of Income and Wealth Distributions} (Cambridge University Press, 2013) ISBN 978-1-107-01344-5

\bibitem{Ribeiro-2020} Marcelo Byrro Ribeiro, \textit{Income Distribution Dynamics of Economic Systems: An Econophysics Approach} (Cambridge University Press, 2020) ISBN 978-1-107-09253-2

\bibitem{Rosser-2016} J. B. Rosser Jr., ``Entropy and econophysics'', Eur. Phys. J. Spec. Top. \textbf{225}, 3091--3104 (2016), \url{http://dx.doi.org/10.1140/epjst/e2016-60166-y}

\bibitem{Shaikh-2017} Anwar Shaikh ``Income Distribution, Econophysics and Piketty'', Review of Political Economy \textbf{29} (2017) 18--29, \url{https://doi.org/10.1080/09538259.2016.1205295}

\bibitem{Scharfenaker-2020} Ellis Scharfenaker, ``Statistical equilibrium methods in analytical political economy'', Journal of Economic Surveys, early view (2020), \url{https://doi.org/10.1111/joes.12403}.

\bibitem{Cho-2014} Adrian Cho, ``Physicists say it's simple'', Science \textbf{344} (2014) 828, \url{https://doi.org/10.1126/science.344.6186.828}

\bibitem{Gronewold-2021} Nathanial Gronewold, ``Anthill Economics: Animal Ecosystems and the Human Economy'', Ch. 3 ``Entropy and Inequality'' (Prometheus Books, 2021) ISBN 978-1-633-88653-7

\bibitem{IRS-1304} Internal Revenue Service (IRS), Statistics of Income (SOI) research division, ``Individual Income Tax Returns", Publication 1304, \url{https://www.irs.gov/statistics/soi-tax-stats-individual-income-tax-returns-publication-1304-complete-report}

\bibitem{Yakovenko-2016} V. M. Yakovenko, ``Monetary economics from econophysics perspective'', Eur. Phys. J. Spec. Top. \textbf{225}, 3313--3335 (2016), \url{http://dx.doi.org/10.1140/epjst/e2016-60213-3}.

\bibitem{Landau-10}  E.~M.~Lifshitz and L. P. Pitaevskii, \textit{Physical Kinetics} (Butterworth-Heinemann, 1981)

\bibitem{Xi-Ding-Wang-2005} Xi, N., Ding, N., and Wang, Y. ``How required reserve ratio affects distribution and velocity of money,'' Physica A \textbf{357} (2005) 543, \url{https://doi.org/10.1016/j.physa.2005.04.014}

\bibitem{Landau-5} L.~D.~Landau and E.~M.~Lifshitz, \textit{Statistical Physics} (Butterworth-Heinemann, 1980)

\bibitem{Lopez-Ruiz-2008} Ricardo L\'opez-Ruiz, Jaime Sa\~{n}udo, and Xavier Calbet, ``A geometrical derivation of the Boltzmann factor'', American Journal of Physics \textbf{76}, 780 (2008), \url{https://doi.org/10.1119/1.2907776}

\bibitem{Biro-2014} Bir\'o TS, V\'an P, Barnaf\"oldi GG, \"Urm\"ossy K., ``Statistical Power Law due to Reservoir Fluctuations and the Universal Thermostat Independence Principle'', Entropy  \textbf{16} (2014) 6497, \url{https://doi.org/10.3390/e16126497}

\bibitem{Lawrence-2013} S. Lawrence, Q. Liu, and V. M. Yakovenko, 
``Global inequality in energy consumption from 1980 to 2010'', 
\textit{Entropy}, \textbf{15}, 5565 (2013), \url{http://dx.doi.org/10.3390/e15125565}

\bibitem{Semieniuk-2020} G. Semieniuk and V. M. Yakovenko, 
``Historical evolution of global inequality in carbon emissions and footprints versus redistributive scenarios'', \textit{Journal of Cleaner Production}, \textbf{264}, 121420 (2020), \url{https://doi.org/10.1016/j.jclepro.2020.121420}

\bibitem{Yakovenko-2001b} A. A. Dr\u{a}gulescu and V. M. Yakovenko, ``Exponential and power-law probability distributions of wealth and income in the United Kingdom and the United States'', Physica A \textbf{299} (2001) 213, \url{https://doi.org/10.1016/S0378-4371(01)00298-9}

\bibitem{Silva-2005} Silva A C and Yakovenko V M ``Temporal evolution of the `thermal' and `superthermal' income classes in the USA during 1983-2001'', Europhys. Lett. \textbf{69}, 304 (2005), \url{https://doi.org/10.1209/epl/i2004-10330-3}

\bibitem{Yakovenko-2001a} A. A. Dr\u{a}gulescu and V. M. Yakovenko, ``Evidence for the exponential distribution of income in the USA'', Eur. Phys. J. B \textbf{20} (2001) 585, \url{https://doi.org/10.1007/PL00011112}

\bibitem{Tao-2019} 
Yong Tao, Xiangjun Wu, Tao Zhou, Weibo Yan, Yanyuxiang Huang, Han Yu, Benedict Mondal, and Victor M. Yakovenko, ``Exponential structure of income inequality: evidence from 67 countries'', Journal of Economic Interaction and Coordination \textbf{14}, 345 (2019), \url{https://doi.org/10.1007/s11403-017-0211-6}

\bibitem{Pareto-1897} Pareto, V., 1897, \textit{Cours d'\'Economie Politique} (L'Universit\'e de Lausanne)

\bibitem{Gibrat-1931} Gibrat R 1931 \textit{Les In\'egalit\'es Economiques} (Sirely, Paris)

\bibitem{Aoki-2003} Fujiwara Y, Souma W, Aoyama H, Kaizoji T and Aoki M ``Growth and fluctuations of personal income'' \emph{Physica A} \textbf{321} (2003) 598, \url{https://doi.org/10.1016/S0378-4371(02)01663-1}

\bibitem{Banerjee-2010} A. Banerjee and V. M. Yakovenko, ``Universal patterns of inequality'', New J. Phys. \textbf{12} (2010) 075032, \url{https://doi.org/10.1088/1367-2630/12/7/075032}

\bibitem{Fiaschi-2012} D. Fiaschi and M. Marsili, ``Distribution of wealth and incomplete markets: Theory and empirical evidence'', Journal of Economic Behavior and Organization \textbf{81} (2012) 243, \url{https://doi.org/10.1016/j.jebo.2011.10.015}

\bibitem{Pearson-1895} Karl Pearson, ``Contribution to the mathematical theory of evolution II: Skew variation in homogeneous material'', Phil. Trans. Royal Soc. London A \textbf{186}, 343 (1895)

\bibitem{Supplement} Supplemental Material is available online in ancillary files on arXiv and with the published paper.

\bibitem{CPI} We used the option ``CPI for All Urban Consumers'', also known as CPI-U, at \url{https://data.bls.gov/cgi-bin/surveymost?bls}

\bibitem{Boghosian-2017}
Bruce M. Boghosian, Adrian Devitt-Lee, Merek Johnson, Jie Li, Jeremy A. Marcq, Hongyan Wang, ``Oligarchy as a phase transition: The effect of wealth-attained advantage in a Fokker–Planck description of asset exchange'', 
Physica A \textbf{476} (2017) 15, \url{https://doi.org/10.1016/j.physa.2017.01.071}

\bibitem{Kakwani-1980} Kakwani N 1980
  \textit{Income Inequality and Poverty} (Oxford University Press, Oxford)

\bibitem{WashPost-2021} Jeff Stein, ``With big tax push, Democrats aim to tackle enormous gains of top 1 percent'', Washington Post, 13 September 2021, \url{https://www.washingtonpost.com/us-policy/2021/09/13/democrats-tax-biden-budget/}


\end{thebibliography}
\end{document}